\documentclass[12pt,english,aps,manuscript]{revtex4}
\usepackage[pdftex]{graphicx}
\usepackage{amssymb,amsfonts,amsmath,verbatim}
\usepackage{epstopdf}
\usepackage{setspace}

\begin{document}
\preprint{Scientific Reports}
\title{Limits of Predictability in Commuting Flows in the Absence of Data for Calibration}

\author{Yingxiang Yang}
\address{
Department of Civil and Environmental Engineering, Massachusetts Institute of Technology, 77 Massachusetts Avenue, Cambridge, MA 02139, USA
}
\author{Carlos Herrera}%
\address{
Department of Applied Math, Universidad Polit\'{e}cnica, Madrid, 28040, Spain
}
\author{Nathan Eagle}
\address{
Department of Epidemiology, Harvard School of Public Health, Boston, MA 02115, USA
}%

\author{Marta C. Gonz\'{a}lez}
\address{
Department of Civil and Environmental Engineering and  Engineering Systems, Massachusetts Institute of Technology, 77 Massachusetts Avenue, Cambridge, MA 02139, USA
}

\begin{abstract}
The estimation of commuting flows at different spatial scales is a fundamental problem for different areas of study. Many current methods rely on parameters requiring calibration from empirical trip volumes. Their values are often not generalizable to cases without calibration data. To solve this problem we develop a statistical expression to calculate commuting trips with a quantitative functional form to estimate the model parameter when empirical trip data is not available. We calculate commuting trip volumes at scales from within a city to an entire country, introducing a scaling parameter $\alpha$ to the recently proposed parameter free radiation model. The model requires only widely available population and facility density distributions. The parameter can be interpreted as the influence of the region scale and the degree of heterogeneity in the facility distribution. We explore in detail the scaling limitations of this problem, namely under which conditions the proposed model can be applied without trip data for calibration. On the other hand, when empirical trip data is available, we show that the proposed model's estimation accuracy is as good as other existing models. We validated the model in different regions in the U.S., then successfully applied it in three different countries.
\end{abstract}
\maketitle
\section*{Introduction}
Good estimates of how many people frequently travel between two places is related to several areas of study. It is not only a key ingredient in modeling the spreading of infectious diseases \cite{balcan2009multiscale, eggo2011spatial, viboud2006synchrony, ramasco2009commuting}, but also a fundamental problem in geo-spatial economics of facility distribution and transportation planning. Clues on the statistics of human movements and their changes at different scales are thus fundamental. Here we explore statistical expressions to calculate the number of commuting trips within areas of sizes ranging from a few kilometers (one city or town) to over one thousand kilometers (one country). We focus on commuting trips because they are stable in time and account for the largest fraction of the total flows in a population.

In order to describe the commuting flow patterns of people, various aggregated trip distribution models  have been proposed \cite{ortuzar1994modelling, zipf1946p, wilson1998land, stouffer1940intervening, wilson1970entropy, wilson1969use}. Among all these efforts, the gravity model, which assumes that the number of movements between an origin-destination pair decays with their distance, is the most widely used one \cite{zipf1946p, erlander1990gravity, choukroun1975general}. The gravity model can be further divided into the unconstrained gravity model \cite{ortuzar1994modelling}, the singly constrained gravity model (such as the B.P.R model \cite{duffus1987reliability, kirby1974theoretical}, or the Voorhees gravity model \cite{erlander1990gravity}) and the doubly constrained gravity model \cite{anas1983discrete}. More constraints generally leads to better model performance. The doubly constrained gravity model fits parameters in an iterative way to best reconstruct the empirical OD matrix. But the iterative fitting complicates the calibration of the distance decaying function, which is essential for the gravity model. However, the most important limitation of the doubly constrained gravity model is the calibrated parameters from the empirical trip data in one area lack meaningful interpretation and thus are not transferable to other regions. As a consequence, in current epidemiology studies the more generalizable but less accurate unconstrained gravity model is often used~\cite{balcan2009multiscale}.

From another perspective, the recently proposed radiation model \cite{simini2012universal, simini2013human} has an analytical formulation that estimates trip volumes without parameters; using as input the distribution of population only. It takes inspiration from the intervening opportunity model, which assumes that the trip volume is more related to the number of opportunities between the origin and the destination instead of just to their distance \cite{stouffer1940intervening, stouffer1960intervening}. As is recently shown by Simini {\it et al.} \cite{simini2013human}, the radiation model represents a general framework that includes the intervening opportunity model as a degenerated case and it is linked with the the gravity model through the spatial distribution of opportunities. The model was originally developed to predict inter-county flows, which not necessarily account for daily trips. Some recent works~\cite{Lenormand2012,Xiao2013,Masucci2013} have shown that the parameter-free radiation model~\cite{simini2012universal} does not work well at predicting intra-city trips. Two of these works ~\cite{Lenormand2012,Xiao2013} introduced models with a calibrating parameter as a cost function of the distance to reproduce intra city trips. In the gravity model based approach~\cite{Lenormand2012} a scaling parameter depending on region size is introduced; while in intervening opportunity model based approaches introducing such an interpretable parameter that can be estimated without trip data remains an open question.

The present work seeks to answer two questions: How the value of the parameter that imposes the scale dependency can be interpreted and estimated without trip data? Under which conditions the prediction of models not calibrated with empirical trip data would work? The extended radiation model that we present in this study is distinct from previous formulations, in that it is a stochastic model that depends on the distributions of opportunities and population only. The one scaling parameter $\alpha$ depends on the region size and the heterogeneity of opportunity distribution, which makes it interpretable and estimable in many cases even without trip data. The extended radiation model is derived by relating the trip production and attraction process to survival analysis. We propose an analogy between survival time $t$ and the number of opportunities $a$ a commuter has considered. Another important ingredient for modelling trips within small scales (intra urban trips) is the separation between population density and trip attraction rates. Most models on intra-city trips use the density of population~\cite{Xiao2013, Masucci2013} as a proxy for both trip generation and trip attraction rates. While this approximation is reasonable at large scales, at inner city scale trip attraction is better represented by the density of point of interests (POIs), defined as geolocated non residential establishments presented on a digital map.

The proposed model combines the closed analytical form of the original radiation model and the flexibility of gravity-like models. While it can be calibrated when empirical trip data is available, it also provides an analytical parameter estimation when there is no trip data for calibration. The model is validated in the U.S. by the census commuting data and in three other countries (Portugal, Dominican Republic, and Rwanda) by cell phone records.

\section*{Results}
\subsection*{Evaluation of the gravity model and the radiation model}
The simplest form of the gravity model is\cite{balcan2009multiscale, eggo2011spatial, viboud2006synchrony, ramasco2009commuting}:
\begin{equation}
T_{ij}=\gamma\frac{n_i^{\eta}n_j^{\beta}}{C(r_{ij})}
\end{equation}
where $T_{ij}$ is the flow between zone $i$ and $j$. $n_i$ and $n_j$ are the population of the two zones. $r_{ij}$ is the distance between them and $C$ is a distance decaying function. $\eta$ and $\beta$ are parameters to be fitted from data. $\gamma$ is an adjustment parameter controlling the sum of the flows. This is usually called the unconstrained gravity model because it does not guarantee the attainment of the desired generation and attraction marginal volumes in each zone.

In transportation planning, the gravity model usually takes the form \cite{anas1983discrete, erlander1990gravity, wilson1974some}:
\begin{equation}
T_{ij}=\frac{\eta_{i}\beta_{j}O_{i}D_{j}}{C(r_{ij})}
\end{equation}
where $O_i$ and $D_j$ are the total trip production and attraction volumes of zones $i$ and $j$ respectively. For a study region with $N$ zones, there are $2N$ parameters of $\eta_i$ and $\beta_j$. These parameters are calculated by iteratively applying: $\eta_i=1/\sum_{j}\beta_jD_jC(r_{ij})$ and $\beta_j=1/\sum_{i}\eta_iO_iC(r_{ij})$. This is called the doubly constrained gravity model because it ensures consistent values of the trip production $\sum_{j}T_{ij}=O_i$, and trip attraction $ \sum_{i}T_{ij}=D_j$ per zone. In order to calibrate the $\eta_i$ and $\beta_j$ parameters, the model requires accurate input of the total trip production and attraction volumes $O_i$ and $D_j$.

An alternative model that does not require calibration is the recently proposed radiation model, in which $T_{ij}$ takes the form:
\begin{equation}
\label{eq:radiation}
T_{ij}=O_i\frac{n_in_j}{(n_i+s_{ij})(n_i+n_j+s_{ij})}
\end{equation}
where $s_{ij}$ is the population within the circle of radius $r_{ij}$ centered at zone $i$ (not including the population in zones $i$ and $j$) and the rest of the notations are the same as in the gravity model.

We explore the suitability of the doubly constrained gravity model and the radiation model on predicting commuting flows at three different scales: the Western U.S., the entire San Francisco Bay area, and the city of San Francisco (see Methods).

\begin{figure*}[b]
\centerline{\includegraphics[width=1\columnwidth]{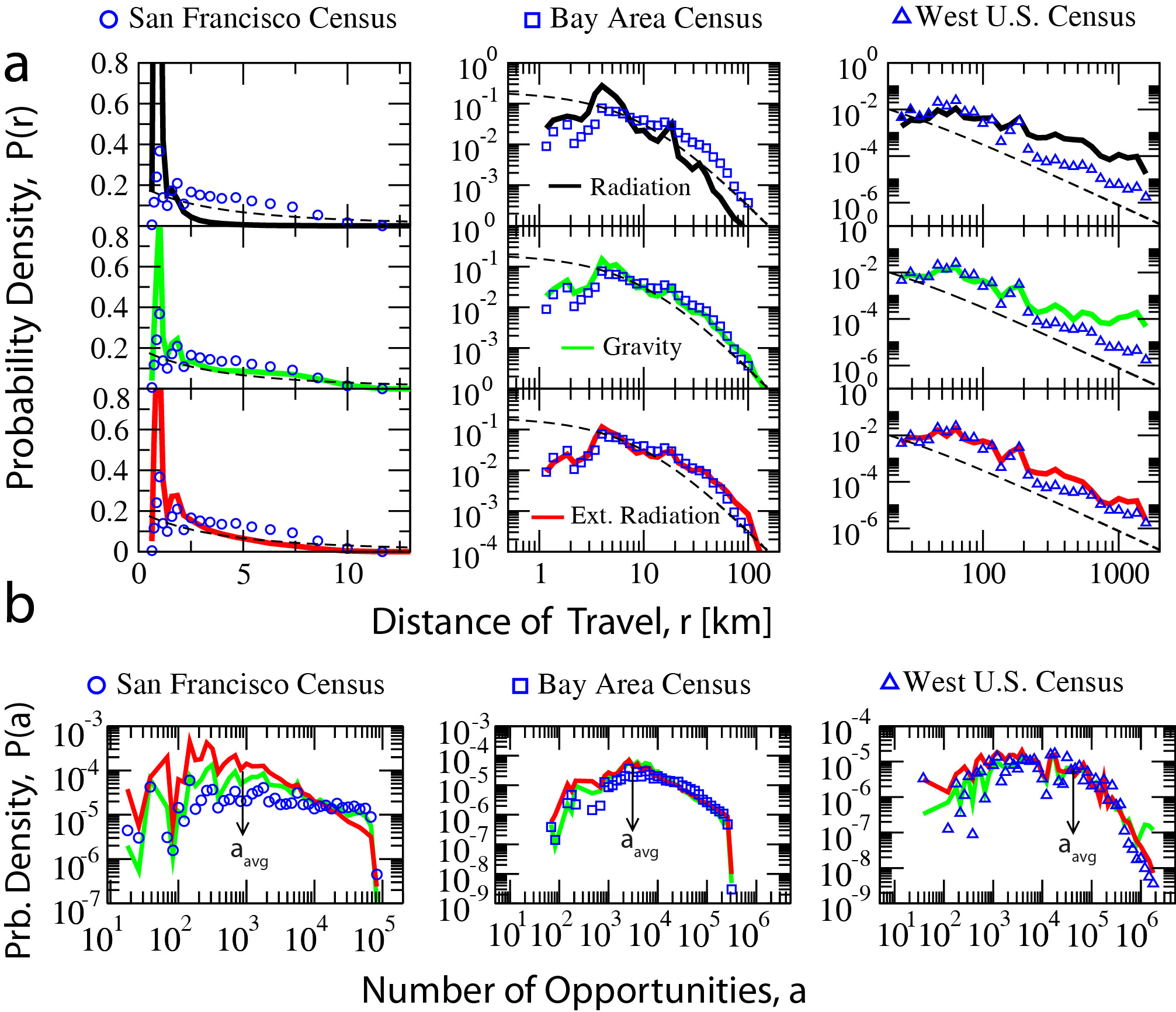}}
\caption{Comparison of the census data, the calibrated doubly constrained gravity model, the parameter-free radiation model, and the calibrated extended radiation model. (a) The three columns represent San Francisco, the Bay area, and the Western U.S. respectively. The three rows are the results of the three different models compared with the census data. The radiation model gives relatively good prediction only at the Western U.S. scale. At the two smaller scales the radiation model under-estimates long distance trips. The doubly constrained gravity model gives close predictions to the census data at the three scales. The extended radiation model, with one parameter $\alpha$, achieves the same prediction quality. The dashed line is a guide to the eye with the distance decaying function $P(r)=100(r+10)^{-2.7}$. (b) The $P(a)$ distribution, is the probability of measuring a commuting trip with $a$ opportunities between the origin and the destination. Because the radiation model is not suitable for the two smaller scales, here only the extended radiation model and the doubly constrained gravity model are compared with the census data. The flat distribution of $P(a)$ in the Census data within San Francisco differs from the other two scales, showing that the distribution of intra-city flows is influenced less by the number of opportunities between the origin and destination.}\label{fig2}
\end{figure*}

We apply the doubly constrained gravity model with power distance decay function $C(r)=r^{k}$ (which is better than the exponential decay function in the example regions) and compare it with the radiation model. Fig.~\ref{fig2}(a) shows the commuting distance distribution $P(r)$ of different models at the three scales of study. When we compare inter-county trips in the Western U.S. both the parameter-free radiation model and the calibrated gravity model with $2N+1$ parameters perform similarly. Adjusting the $\eta_i$, $\beta_j$ and the parameter $k$ in the distance decaying function can not fit the model well for both short distance trips and long distance trips. This confirms the results reported in~\cite{simini2012universal}.

However, when trying to predict the commuting flows among zones within the Bay area or San Francisco, without parameters for calibration the situation is much harder because the density of population is more homogeneously distributed and commuters tend to go to various business districts across the area (see Fig.~\ref{fig1}). Such scale of daily trips is where the calibration parameters start playing an important role and the calibrated gravity model performs better than the parameter-free radiation model.
In order to inspect further this situation, the distribution of the total number of opportunities $a$ between trip origin and destination is calculated. Fig.~\ref{fig2}(b) shows that at the Bay area scale (zone size $l\approx10 km$), there is a region $a<a_{avg}$ where there is not a clear functional form on the enclosing number of opportunities $a$ between the origin and the destination ($a_{avg}$ is the average number of opportunities in a zone). While for $a>a_{avg}$ the probability of finding a trip start monotonically decaying. This effect of clear decaying behavior for $a>a_{avg}$ is not observed in commuting trips within San Francisco. Based on these observations we look for a way to introduce the effects of scale on the radiation model.

\subsection*{Extension of the radiation model}
We introduce a scaling parameter $\alpha$ by combining the derivation of the original radiation model with survival analysis. Survival analysis uses statistical methods to deal with the analysis of time to events, such as life time distribution of living organisms or machine components \cite{miller2011survival}. The two objects of primary interest are the survival function and the hazard function. The survival function, $S(t)$, represents the life time distribution of an entity:
\begin{equation}
  S(t)=Pr(T>t)
\end{equation}
$S(t)$ is the cumulative probability of having survived after time $t$. The hazard function, $h(t)$, represents the conditional death rate at time $t$:
\begin{equation}
\begin{aligned}
  h(t)&=\lim_{dt\rightarrow 0}\frac{Pr(t\leq T<t+dt\mid T\geq t)}{dt} \\
\end{aligned}
\end{equation}
Different forms of hazard function $h(t)$ will lead to different survival function $S(t)$. The two most generally used hazard functional forms are $h(t)=\lambda$ and $h(t)=\lambda \alpha t^{\alpha-1}$. $h(t)=\lambda$ leads to the exponential survival function $S(t)=e^{-\lambda t}$, while $h(t)=\lambda \alpha t^{\alpha-1}$ leads to the Weibull survival function $S(t)=e^{-\lambda t^\alpha}$. For $\alpha \rightarrow 0$ the hazard function does not depend on time $t$, this is the effect that we would like to replicate for smaller zones: job selection has low dependence on the number of opportunities between home and the selected work location.

The intervening opportunity model can be derived under the survival analysis framework if we think of the departure from origin as 'birth' and the arrival at the destination as 'death' while the lifetime is measured as the number of opportunities, $a$, between the origin and the destination. The survival function $S(a)$ represents the cumulative probability of not finding a workplace within $a$ opportunities. $S(a)=e^{-\lambda a}$ and $S(a)=e^{-\lambda a^\alpha}$ are exactly the two most commonly used intervening opportunity models \cite{stouffer1940intervening, stouffer1960intervening}. Note here that $\alpha$ controls the slope of the decay of $S(a)$.

We show here the radiation model can also be derived under the survival analysis framework. If $P_>(a)$ represents the probability of not accepting the closest $a$ opportunities, it has the same meaning of the survival function $S(t)$ in the survival analysis. A person commuting from a region of $n_i$ opportunities to a region of $n_j$ opportunities with $s$ opportunities in between, can be expressed as the conditional probability of accepting one of the $n_j$ opportunities between $a$ and $a+n_j$, given that the closest $n_i$ opportunities are not chosen. Note that $a\equiv n_i+s$ for Eq. \ref{eq:radiation}. This probability can be written as:
\begin{equation}\label{eq:p1mna}
  P(1|n_i,n_j,a)=\frac{P_>(a)-P_>(a+n_j)}{P_>{(n_i)}}
\end{equation}
Then the core question becomes how to get the expression of $P_>(a)$. Different individuals should have different hazard rates (or job expectations in this case) \cite{harrell2001regression,hosmer2011applied}, we can assign different parameter $\lambda_i$ value to different individuals $i$. So the survival function of the entire population from a given origin becomes:
\begin{equation}\label{eq:sa}
  P_>(a)=E[e^{-\lambda_i a}]=\int_{0}^{+\infty}e^{-\lambda a}p(\lambda)d\lambda
\end{equation}
 If we define $p(\lambda)=e^{-\lambda}$ and $\lambda\in(0,+\infty)\rightarrow e^{-\lambda}\in(0,1)$:
\begin{equation}
\begin{aligned}
  P_>(a)=\int_{0}^{+\infty}e^{-\lambda a}p(\lambda)d\lambda=\int_{0}^{+\infty}e^{-\lambda a}e^{-\lambda}d\lambda=\frac{1}{1+a}
\label{eq:pa1}
\end{aligned}
\end{equation}
which gives the radiation model expression introduced in Eq.\ref{eq:p1mna} (we are using the same notation as in \cite{simini2013human}). One reason of choosing $p(\lambda)$ to be an exponential distribution is it connects the survival analysis with the derivation of the original radiation model \cite{simini2013human}. We further elaborate the influence of the form of the $p(\lambda)$ distribution in the SI Appendix. If we extend the analysis to the Weibull survival function, we get:
\begin{equation}
\begin{aligned}
\label{eq:pa}
    P_>(a)=\int_{0}^{+\infty}e^{-\lambda a^{\alpha}}p(\lambda)d\lambda=\frac{1}{1+a^{\alpha}}
\end{aligned}
\end{equation}
Eq. \ref{eq:p1mna} becomes:
\begin{equation}\label{p1mnafull}
\begin{aligned}
  P(1|n_i,n_j,a_{ij})=\frac{[(a_{ij}+n_j)^\alpha-a_{ij}^\alpha](n_i^\alpha+1)}{(a_{ij}^\alpha+1)[(a_{ij}+n_j)^\alpha+1]}
\end{aligned}
\end{equation}
In order to calculate the flows between two zones $i$ and $j$, two normalization constants are needed:
\begin{equation}\label{tij}
  T_{ij}=\gamma m_i\frac{P(1|n_i,n_j,a_{ij})}{\sum_k P(1|n_i,n_k,a_{ik})},
\end{equation}
since not all people are commuting and also the commuting within each zone is not counted, $\gamma$ is the percentage of population that is commuting between different zones in the study area. $m_i$ is the population at the origin. If the empirical trip data or cell phone records (as will be shown in the sections below) are available, $\gamma$ can be calculated from the total number of observed trips. If neither data source is available, the flow distribution can still be calculated because $\gamma$ doesn't influence the relative ratio of flows between different OD pairs. The denominator in the equation is a normalization constant for finite sized area. The influence of the border effect and how this formulation can solve some of the limitations in previous models is detailed in the SI Appendix.

The extended radiation model is calibrated to the three regions examined in the previous section. The obtained $\alpha$ values are $0.003$, $0.05$ and $1.5$ respectively. We use the common part of commuters (CPC), based on the S\o rensen index \cite{sorensen1948method}, to quantitatively measure the goodness of flow estimation.
\begin{equation}\label{cpc}
CPC(T,\widetilde{T})=\frac{2NCC(T,\widetilde{T})}{NC(T)+NC(\widetilde{T})}
\end{equation}
$NCC(T,\widetilde{T})=\sum_{i=1}^{n}\sum_{j=1}^{n}\textup{min}(T_{ij},\widetilde{T_{ij}})$, $NC(T)=\sum_{i=1}^{n}\sum_{j=1}^{n}T_{ij}$. This index shows which part of the commuting flow is correctly estimated, 0 means no agreement found and 1 means perfect estimation. Table~\ref{table1} shows that the extended radiation model gives estimations with similar performance to the doubly constrained gravity model at the three regions while the original radiation model's estimation power decays with the region granularity. The goodness of fit of the extended radiation model is close to other recently proposed models~\cite{Lenormand2012}. The difference is that the study in~\cite{Lenormand2012} uses actual commuting flow generation and attraction volumes as input, while in the current model we use more easily acquired population and POI density as proxies, but achieved the same level of accuracy.
\subsection*{Absence of data to calibrate $\alpha$}
The $P_>(a)$ distribution plays an important role in the formulation and $\alpha$ parameter calibration, thus worths further scrutiny. When the space is infinite and the opportunities are continuous $P_>(a)=\frac{1}{1+a^\alpha}$ is a monotonically decreasing function with slope given by the $\alpha$ value. But if we are considering trips only within a finite sized region, this implies a finite numbers of opportunities possible, up to $a_{tot}$. Thus $P_>(a_{tot})=0$. Also, we divide a study region into a finite number of zones $n_{cells}$, so $a$ can only take a set of discrete values. We define $a_{avg}\equiv a_{tot}/n_{cells}$. $a_{min}$ is the smallest number of opportunities in all the zones. Since within zone trips are not considered, $P_>(a)$ should start to decrease after $a_{min}$. This value is not known a priory but may be approximated by $a_{avg}$ in the absence of data on trips. We correct the expression in Eq. \ref{eq:pa} to account for these effects as:
\begin{equation}\label{patheo}
\langle P_>(a) \rangle =\frac{\frac{1}{1+a^\alpha}-\frac{1}{1+a_{tot}^\alpha}}{\frac{1}{1+a_{avg}^\alpha}-\frac{1}{1+a_{tot}^\alpha}}, a_{tot}\geq a \geq a_{avg}
\end{equation}
Now, we explore how Eq.~\ref{patheo} can reproduce the $P_>(a)$ measured from data. The top panel of Fig. 3(a) shows the results of $P_>(a)$ calculated from the census commuting data in San Francisco, the Bay area and the Western U.S. The solid lines show Eq.~\ref{patheo} with  different $\alpha$ values, note that the two limiting values of $a_{avg}<a<a_{tot}$ determine  the range of the equation. By comparing Eq.~\ref{patheo} with the data as seen in Fig.~\ref{fig2}(b) and Fig.~\ref{fig3}(a), we see that in the intra-city scale the $P_>(a)$ distribution does not decay beyond $a_{avg}$, so we can't use Eq.~\ref{patheo} to estimate the radiation model parameter. For the Bay area and the Western U.S. scale, $a_{min} ~\sim a_{avg}$, Eq.~\ref{patheo} works well and the value of $\alpha$ ($0.1 \le \alpha \le 2$) should increase with the scale $l$. For a fixed scale $l$, if the heterogeneity of opportunity distribution increases, then $a_{min}$ further differs from $a_{avg}$. In those cases $\alpha>2$ and it is not possible to estimate $\alpha$ without data calibration (as shown for Las-Vegas-Seattle in Fig. 3(a)). The detailed explanation is in the rest of this section.

We further explore how the parameter $\alpha$ systematically changes as the size $l$ of the commuting zones changes. We evaluate the commuting within regions divided into $n_{cells}=100$ zones of size $l$ ranging from a few kilometers to over 100 kilometers (See Fig. 3 and Fig. 4). We randomly chose $200$ study regions for each scale with total population of at least $5,000 \times l$ in order to avoid unpopulated regions such as national parks. The census commuting OD data is used to calibrate the $\alpha$ value in each region using Eq. \ref{tij}. Fig. 3(b) shows how $\alpha$ is close to zero for $l<10$ $km$ and starts to increase as a power function beyond that value. The functional relationship as a solid line is:
\begin{equation}
\label{eq:la}
 \alpha=(\frac{l}{36[km]})^{1.33}
\end{equation}
The error bars show the 20th and 80th percentile of the $\alpha$ value at each scale. The three cases calibrated before: San Francisco, the Bay area, and the Western U.S. are marked in red squares. They are all close to the expected values calculated from Eq.~\ref{eq:la}. For trips within a city and up to metropolitan urban areas the $\alpha$ value is close to $0$ and the error bar is narrow. In this limit ($\alpha\rightarrow 0$) we have:
\begin{equation}
\begin{aligned}
P(1|n_i,n_j,a_{ij})&=\lim_{\alpha\rightarrow 0}\frac{[(a_{ij}+n_j)^\alpha-a_{ij}^\alpha](n_i^\alpha+1)}{(a_{ij}^\alpha+1)[(a_{ij}+n_j)^\alpha+1]}\\
&=\lim_{\alpha\rightarrow 0}\frac{(a_{ij}+n_j)^\alpha-a_{ij}^\alpha}{2}
\end{aligned}
\end{equation}
The ratio between $(a_{ij}+n_j)^\alpha-a_{ij}^\alpha$ and $a_{ij}^\alpha$ is:
\begin{equation}
\begin{aligned}
  \lim_{\alpha\rightarrow 0}\frac{(a_{ij}+n_j)^\alpha-a_{ij}^\alpha}{a_{ij}^\alpha}=\lim_{\alpha\rightarrow 0}\alpha\frac{n_j}{a_{ij}}
\end{aligned}
\end{equation}
The $\alpha$ will cancel out when substituting the expression back to Eq.~\ref{tij}. The detailed derivation is in the SI Appendix. This implies that the predicted flow is proportional to $\frac{n_j}{a_{ij}}$, while in the parameter-free radiation model, the flow is proportional to $\frac{n_j}{a_{ij}^2}$. We have mentioned that within such short distances Eq.~\ref{patheo} is not suitable for the exact estimation of $\alpha$. But given that at such scale $0<\alpha<0.1$, and the model is not sensitive to the parameter value when $\alpha\rightarrow 0$, in zones that $l<10km$ Eq.~\ref{patheo} is able to generate a reasonable approximated $\alpha$ value for trip volume estimation.

In the range $l \sim 10...65$ km, $0.1\leq \alpha \leq 2$ most commonly accounts for inter city trips, the model without data calibration is expected to predict trips well because of the narrow error bar. For larger scale regions enclosing trips between two or more combined statistical areas such as the ones shown in Fig. 4, $\alpha>2$ and has a wide range. We notice that the main driving factor influencing the $\alpha$ value for the same scale is the differences in the homogeneity of facility density; which are highly correlated to population density at these large scales.

In Fig. 4, the two marked regions have the same scale: $l=60$ km. The population distribution of the southern region has two sharp centers, Los Angeles and Las Vegas, while the rest has low population density. In the northern region, the population is more homogeneously distributed. One example OD pair is shown for each region on the right part of Fig. 4: From Los Angeles to Lake Havasu City for the southern region and from Seattle to Wenatchee for the northern region. They have similar $m_i$, $n_i$, $n_j$ and $s_{ij}$ values. According to the original radiation model, they should have similar flow volumes. But in the census there are only $26$ people commuting from Los Angeles to Lake Havasu City while there are $167$ commuting from Seattle to Wenatchee. The reason is quite clear on the map: the distance from Seattle to Wenatchee is only $150$ km while the distance from Los Angeles to Lake Havasu City is much longer because of the low population/opportunity density between the origin and the destination. To put it in another way, people have to travel further to be able to explore the same amount of opportunities. This causes the calibrated $\alpha$ value of the southern region to be $5$, much larger than the northern one, which is $1.6$. As is shown in Fig. 3(a), the more heterogeneous the distribution of population is; the larger the difference in $a_{min}$ and $a_{avg}$ is and the larger the $\alpha$ value becomes. In those cases the $P_>(a)$ from empirical trip data differs more from Eq.~\ref{patheo} and empirical trip data is needed for parameter calibration. More quantitative ways to measure the influence of the degree of heterogeneity as a cost function depending on the distance between the origin and the destination remains an open question for further studies.

People's location choices are not influenced by the choice of study region sizes. What the parameter $\alpha$ captures for the scale dependency is the granularity of aggregation. Ideally the location choice should be modeled to the smallest spatial granularity, then aggregated to the desired granularity level. But in practice such fine grained input data are usually not available, in such cases the $\alpha$ parameter helps the model estimation at the desired granularity directly, without requiring finer grained data.

In summary, even without empirical trip OD data, if the commuting zones are in the range $l \sim 1...65 km$, it can be expected for the extended radiation formula to give good commuting flow prediction. In these cases Eq.~\ref{eq:la} gives us: $0\leq \alpha \leq 2$. For zones with large sizes, if the opportunity distribution do not have strong heterogeneity ($a_{min} \sim a_{avg}$), the monotonic increase of $\alpha$ with scale $l$ is usually captured by Eq.~\ref{patheo}. In other situations the model needs to be calibrated with empirical OD data.
\subsection*{Multi-regional study and the role of cell phone data}
Not many countries in the world have detailed census data for commuting flow prediction and model calibration. Those countries with data scarcity are often developing countries that need this kind of modeling the most. For these countries finding an alternative data source to provide guidance for their urban growth, economic planning and epidemics controlling is a pressing need. In this section we show how the extended radiation and the gravity model can be calibrated given estimated commuting trips measured from cell phone data. More importantly we can compare the phone data calibrated parameter $\alpha$ with the one predicted from Eq. ~\ref{eq:la} to explore the generality of that expression.

Cell phone records are increasingly showing the potential to become a data source of valuable information \cite{lu2012predictability, wesolowski2012quantifying, bagrow2011collective} since most populated areas have cell phone service coverage and the value of cell phone data in modeling human mobility has recently been highlighted in various studies \cite{gonzalez2008understanding, song2010limits, song2010modelling, balcan2009multiscale}. For instance, in Rwanda there is no detailed commuting census data available. Even if there were, the high migration rate of people would make the census outdated quickly. However, the country has $215,030,420$ cell phone records from one cell phone service provider in just three months. In this section cell phone records are used to extract a seed commuting OD matrix, which is expanded using iterative proportional fitting to estimate the full commuting OD matrix for the whole population under study.

We use the Bay area as an example to validate the method. Fig. 5(a-c) shows the comparison of the results of the distribution of commuting distance, the distribution of the number of commuters between O-D pairs, and the comparison of the census commuting flow $T_{ij}$ with the expanded cell phone user commuting flow $T'_{ij}$. The close fitting in all the three figures shows that we can recover the commuting patterns of the whole population from the seed matrix provided by cell phone records. For countries that do not have population density census statistics for the IPF expansion, we can use the Landscan \cite{bhaduri2007landscan} population density estimation which is available worldwide at $1km^2$ resolution.

We then extended our study to three different countries: Portugal, Dominican Republic and Rwanda. We selected the capitals in the three countries: Lisbon, Santo Domingo (including the greater metropolitan area), and Kigali; and also did the analysis for the entire Rwanda and Portugal (we do not have the cell phone information available for the entire Dominican Republic). We calibrated the gravity model and the extended radiation model to test how much can they recover these regions' commuting patterns. The results are shown in Fig. 5(d-h). The difference in the commuting distance distribution in these regions are captured by both models. The values of $l$ and $\alpha$ for the extended radiation model of these regions are marked as triangles in Fig. 3(b). All of them conform to the functional form of $\alpha$ observed from the U.S. regions. This shows that the relationship between $\alpha$ and the scale $l$ is generalizable, in this case we could have used the extended radiation formula in these countries to estimate trips, in the absence of trip data to calibrate the model.

\section*{Discussion}
In summary, we propose an extension to the radiation model that can be calibrated with one scale parameter to predict commuting flows at different spatial scales. The scale parameter $\alpha$ modulates the influence of the opportunity distribution heterogeneity and the spatial scale $l$ of the commuting zones.

Our results are then compared with the benchmark model calibrated with trip data, known as the doubly constrained gravity model. A multi-scale study shows that both the extended radiation model and the doubly constrained gravity model give close estimations to the census commuting flows.

The main advantage of the proposed modelling framework is that it still can be applied to predict number of commuting trips when lacking data for calibration. The $\alpha$ parameter depends on the scale of the study region and the homogeneity of the population distribution. We show that for Eqs. \ref{p1mnafull} and \ref{tij} to give good results, the size of the $n_{cells}$ commuting zones is in the range $l \sim 1...65$ km, representing $0\leq \alpha \leq 2$. These values of $\alpha$ imply mild heterogeneity in the distribution of opportunities among zones; which here means that the minimum number of opportunities $a_{min}$ enclosing trips is close to $a_{avg}$ (the average number of opportunities for all the zones, $a_{avg} \equiv a_{tot}/n_{cells}$). Other quantitative ways to measure the degree of heterogeneity remains an open question for further studies. For larger regions ($l>65$ km) the $\alpha$ value range is wide because the heterogeneity of population distribution is highly variable at this scale. In these cases the model is better used with empirical data for parameter calibration rather than estimating $\alpha$ from Eq.~\ref{eq:la} directly.

The presented study provides the first building blocks for a multi-scale generator of human mobility expressed as a functional form of the distributions of population and job facilities. We tested it in different scales at different countries and discussed its range of applicability. We share the sample of the U.S. county level commuting flow prediction on our web-page to help in this direction~\cite{webpage}.
\section*{Methods}
\subsection*{Input data preparation}
\begin{figure*}[t]
\centerline{\includegraphics[width=1\columnwidth]{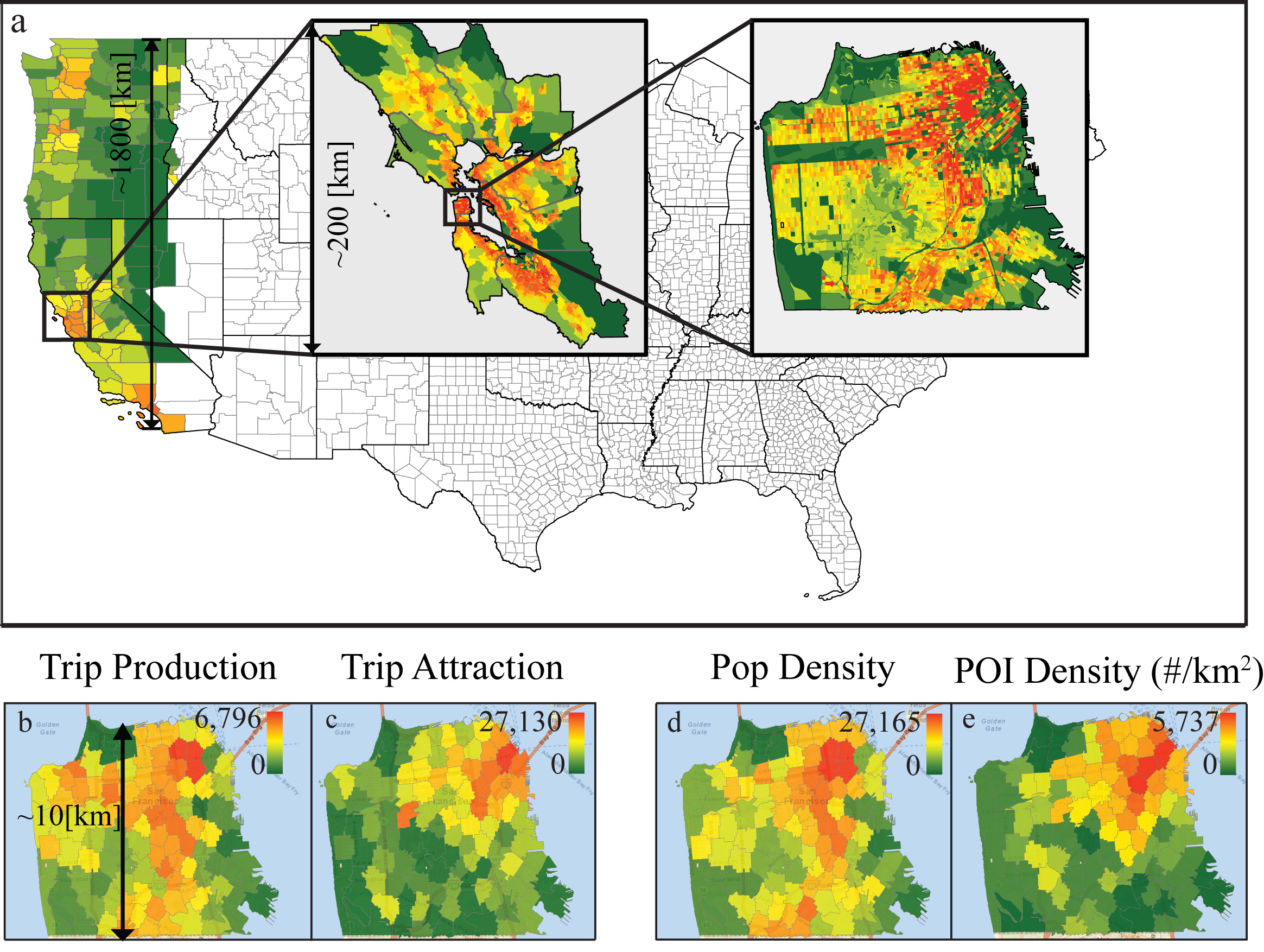}}
\caption{Trip production and attraction at different scales. (a) A map showing the three selected regions of study representing: the Western U.S., the Bay area, and San Francisco. The color bar represents population density. (b-e) Commuting trip generation rate, trip attraction rate, the population density and the POI density in ($\#/km^2$) in San Francisco. While the population density has high correlation with the commuting trip generation rate, the POI density has high correlation with the commuting trip attraction rate. In contrast, at the scale of the Western U.S., the population density correlates with both the trip generation rate and the attraction rate, as shown in Table~\ref{table2}. The maps are generated from ARCGIS using TIGER/Line Shapefiles.}\label{fig1}
\end{figure*}

\begin{figure*}
\centerline{\includegraphics[width=1\columnwidth]{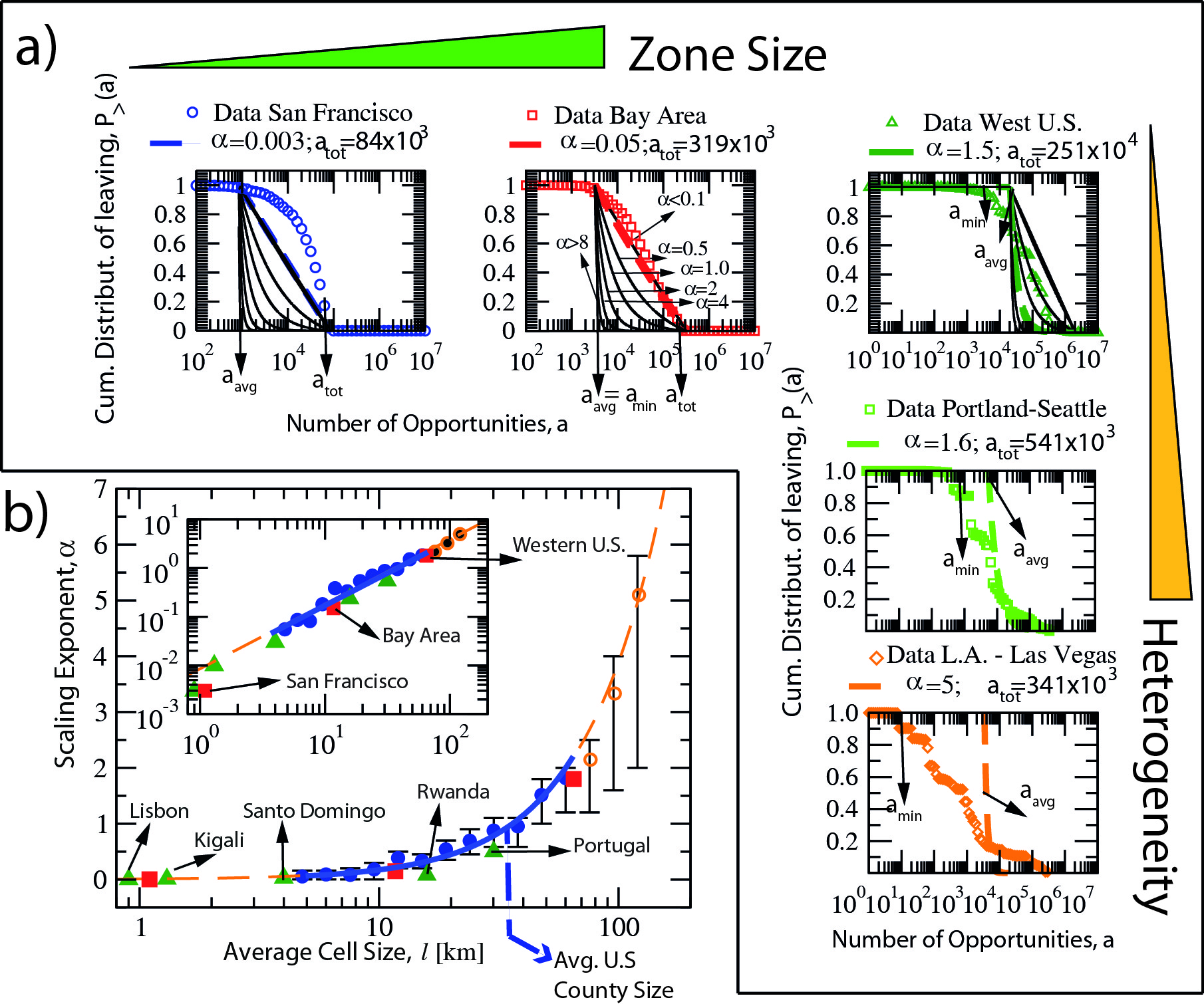}}
\caption{Interpretation of the parameter $\alpha$. (a) Effects of the scale on the $P_>(a)$ distribution, showing as symbols the results from census data of San Francisco, the Bay area and the Western U.S. The black solid lines are evaluations of Eq. \ref{patheo} for different values of $\alpha$, and the dashed line is the expected $\alpha$ value using Eq. \ref{eq:la}. The analytical predictions work for the Bay area and Western U.S. From top to bottom we see the $P_>(a)$ distribution for three regions with similar sizes. Given a fixed scale, the $\alpha$ value is influenced by the heterogeneity of the distribution of opportunities. The more heterogeneous the region is, the larger the difference between $a_{avg}$ and $a_{min}$, as is shown in Las Vegas-L.A. region. In these cases the prediction of $\alpha$ (Eq. \ref{patheo}) will not resemble their calibrated values and thus calibration is needed. (b) For each zone scale $l$, $200$ regions with random centers are selected within west U.S. In each case the corresponding $\alpha$ value is calibrated with census trip data. The functional relationship between $\alpha$ and $l$ is $\alpha=(\frac{l}{36})^{1.33}$. The error bar shows the $20$ and $80$ percentile $\alpha$ value for each scale. The inset shows the same plot in logarithmic scale. Marked as solid blue circles is the scale range that $\alpha$ values can be predicted without data calibration. The calibrated results with trip data for U.S. regions are marked as red squares, while the examples from other countries are marked as green triangles. They all follow the functional approximation.} \label{fig3}
\end{figure*}

The three regions, San Francisco, San Francisco Bay Area, and the Western U.S., are shown in Fig.~\ref{fig1}(a). The Western U.S. is divided into $183$ counties while the two smaller regions are divided into $n_{cells}=100$  zones to calculate the ODs. Each zone is a cluster of blocks determined by applying k-means clustering method on the $7,348$ census blocks in San Francisco and $117,219$ blocks in the Bay area. Note that the unconstrained gravity model is not compared here because when there is empirical OD for parameter calibration, the doubly constrained gravity model performs much better. Detailed comparisons between the two gravity models are in the SI Appendix.

We first test the usual assumption in both models: the population density could represent both the commuting trip generation and attraction rates at different scales. We use the $2010$ census LEHD Origin-Destination Employment Statistics (LODES) \cite{LEHD}, which provides home and employment locations for the entire U.S. population at block level. The first column in Table~\ref{table2} shows the correlations between densities ($\#/km^{2}$) of commuting flow generation, attraction and population in the Western U.S. Both of them have high correlations, so at this scale the assumption holds. Fig.~\ref{fig1}(b, c) shows the commuting trip generation and attraction rates in San Francisco. Their distributions are less similar. Thus, we need to find a better proxy for commuting trip attraction rates at smaller scales.

Digital traces of facilities are available on-line, they provide good estimates of commuting trip attractions \cite{Rodrigues2013,sui2012crowdsourcing, gould2008next}. In this study we use the density of point of interests (in $\#/km^{2}$) of each zone to represent the commuting trip attraction rate($\#/km^{2}$). The three study regions contain $1,774,154$; $319,170$ and $85,230$ POIs extracted from Google Places respectively. According to Table~\ref{table2}, at all the scales POI density has high correlation with the commuting trip attraction rate. A more detailed multi-scale correlation analysis is in the SI Appendix.

\subsection*{Seed cell phone OD matrix expansion}
Each cell phone record has a time stamp and a corresponding cell phone tower. For each user, the most frequently used tower between 6PM and 6AM is assigned as the home location and the most frequently used tower during day is assigned as the work location. Using the 2010 census home and employment location data as a benchmark, we chose the Bay area as an example to validate that the cell phone data could provide accurate predictions to commuting flow patterns. The sample includes $189,621$ cell phone users. We mapped the $892$ cell phone towers in the Bay area to the previously defined $100$ block clusters to get the commuting OD matrix for the cell phone users. The iterative proportional fitting (IPF) method is performed to expand the cell phone user OD matrix to the OD matrix for the entire population \cite{fienberg1970iterative}. The basic procedure is first getting the distribution of population and POIs to represent the marginal distributions of commuting trip generation and attraction rates for each block cluster. Then iteratively adjusting the elements of the seed matrix to let them match the desired margins.

\begin{acknowledgments}
This work was funded by the Austrian Institute of Technology, New England UTC Year 23 grant, awards from NEC Corporation Fund, the Solomon Buchsbaum Research Fund, BMW and France Telecom. The authors would like to thank Dietmar Baure, Jameson Toole, Sarah Chung and Yan Ji for useful comments and technical support.
\end{acknowledgments}

\section*{Author Contributions}
YY and MG designed research; YY and CH performed research; YY, NE and MG wrote the paper.

\section*{Additional information}
The authors declare no competing financial interests.






%

\begin{figure*}
\centerline{\includegraphics[width=1\columnwidth]{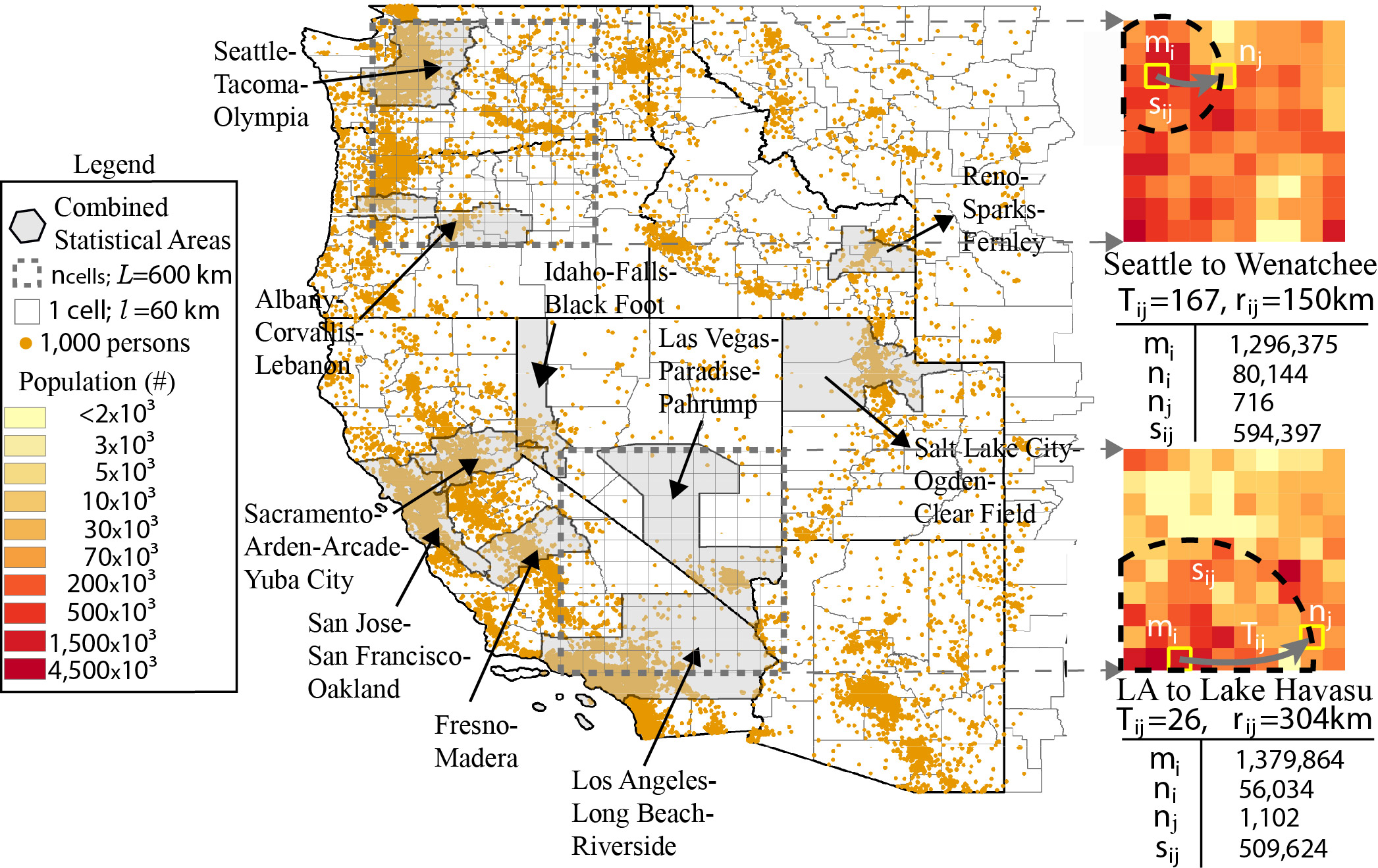}}
\caption{The influence of the opportunity homogeneity on the parameter $\alpha$ value. The two grey rectangles are of the same scale: 600km. The population distribution of the southern region has only two sharp peaks: Los Angeles and Las Vegas, while the population in the northern region is more homogeneously distributed. The right section of the figure shows one example OD pair in each region with similar $m_i$, $n_i$, $n_j$ and $s_{ij}$ values. But the distance between Los Angeles and Lake Havasu City is much longer than the distance between Seattle and Wenetchee, which makes its commuting flow volume much smaller. This effect of distance is taken into consideration by the difference in the $\alpha$ value. For the southern region $\alpha=5$ while for the northern one $\alpha=1.6$. The grey regions are combined statistical areas which usually include one or more populated metropolitan areas.}\label{fig4}
\end{figure*}

\begin{figure*}
\centerline{\includegraphics[width=1\columnwidth]{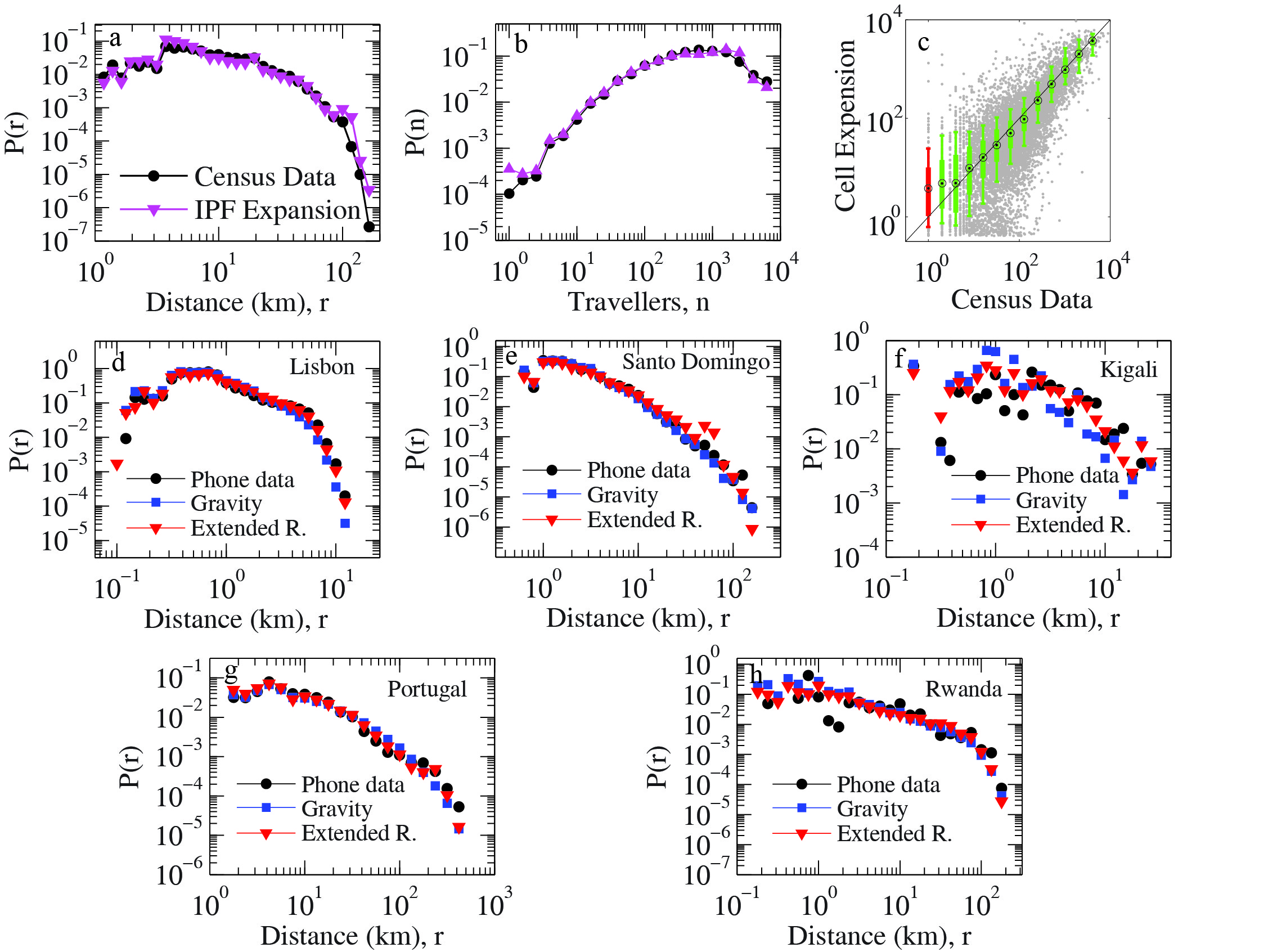}}
\caption{Validation of the IPF expression method and commuting patterns in different regions. (a-c) The comparison of the distribution of commuting distance, the distribution of the number of commuters between O-D pairs, and the comparison of the census commuting flow $T_{ij}$ with the expanded cell phone users' commuting flow $T_{ij}'$. The close fitting in all three figures shows that we can recover the commuting patterns of the whole population from the seed matrix provided by the cell phone records. (d-h) Comparison of the distributions of commuting distance for Lisbon, Santo Domingo, Kigali, Portugal, and Rwanda. The extended radiation model can be successfully applied to all these cases and the corresponding $\alpha$ versus $l$ relationship is marked in Fig.~\ref{fig3}.}\label{fig5}
\end{figure*}

\clearpage
\begin{table}[h]
\centering
\caption{CPC values for different models at different regions}\label{table1}
\begin{tabular}{cccc}
&Western U.S.&Bay area&San Francisco\\
\hline
Ext. Radiation&0.51&0.67&0.65\\
Gravity&0.5&0.64&0.66\\
Radiation&0.43&0.4&0.23\\
\hline
\end{tabular}
\end{table}

\begin{table}[h]
\centering
\caption{Correlation between commuting trip generation, attraction, population and POI density}\label{table2}
\begin{tabular}{ccccccc}
\multicolumn1c{}&\multicolumn2c{Western U.S.}&\multicolumn2c{Bay area}&\multicolumn2c{San Francisco}\cr
\hline
\multicolumn1c{}&\multicolumn1c{Population}&\multicolumn1c{POI}&\multicolumn1c{Population}&\multicolumn1c{POI}&\multicolumn1c{Population}&\multicolumn1c{POI}\cr
\hline
Trip Generation&0.993&0.926&0.971&0.491&0.956&0.292\\
Trip Attraction&0.989&0.930&0.417&0.859&0.157&0.880\\
\hline
\end{tabular}
\end{table}






\end{document}